\documentclass{article}
\usepackage{style/spconf,amsmath,graphicx}
\usepackage{xcolor}
\usepackage{booktabs}
\usepackage{multirow}
\usepackage{subcaption}
\usepackage{enumitem}

\title{S3T: Self-Supervised Pre-training with Swin Transformer \\ for Music Classification}

\name{Hang Zhao$^{1}$, Chen Zhang$^{2}$, Bilei Zhu$^{1}$, Zejun Ma$^{1}$, Kejun Zhang$^{2}$}
\address{$^{1}$ByteDance AI Lab Speech \& Audio Team, $^{2}$Zhejiang University, China}

\begin{document}
%
\maketitle
\begin{abstract}
In this paper, we propose S3T, a self-supervised pre-training method with Swin Transformer for music classification, aiming to learn meaningful music representations from massive easily accessible unlabeled music data. S3T introduces a momentum-based paradigm, MoCo, with Swin Transformer as its feature extractor to music time-frequency domain. For better music representations learning, S3T contributes a music data augmentation pipeline and two specially designed pre-processors. To our knowledge, S3T is the first method combining the Swin Transformer with a self-supervised learning method for music classification. We evaluate S3T on music genre classification and music tagging tasks with linear classifiers trained on learned representations. Experimental results show that S3T outperforms the previous self-supervised method (CLMR) by 12.5 percents top-1 accuracy and 4.8 percents PR-AUC on two tasks respectively, and also surpasses the task-specific state-of-the-art supervised methods. Besides, S3T shows advances in label efficiency using only 10\% labeled data exceeding CLMR on both tasks with 100\% labeled data.

\end{abstract}
\begin{keywords}
Self-supervised learning, Swin Transformer, music genre classification, music tagging.
\end{keywords}

\section{Introduction}
\label{sec:intro}
Automatic music classification categorizes a music piece into labels based on several factors such as genre, mood, artist, and instrumentation. It is essential in the field of music information retrieval (MIR), with a variety of applications such as music recommendation, music retrieval, and music discovery. Therefore, music classification is of interest to both industry and academia and has been actively explored for decades.

Supervised learning methods have dominated the previous music classification works. 
However, these methods normally consume a considerable amount of parallel data to train a model, and collecting paired data requires music annotation with certain expertise, which incurs a high cost. In contrast, unlabeled music data is easy to collect and available in large quantities.
To exploit massive unpaired data, self-supervised learning (SSL) is proposed in recent years and performs well in many fields. SSL conducts task-agnostic pre-training on large amount of unlabeled data and learns representations via designing either heuristics-driven pretext tasks~\cite{oord2018representation,wu2021multi} or augmentation-invariant contrastive methods~\cite{he2020momentum,chen2020simple}, then the learned representations are transferred to a series of downstream tasks.

Recently, several works have made attempts to improve music classification performance with self-supervised learning.
CLMR~\cite{spijkervet2021contrastive} applied an SSL framework SimCLR~\cite{chen2020simple} with SampleCNN~\cite{lee2018samplecnn} as encoder on raw waveforms. SimCLR benefits from large batch size due to its in-batch contrast setting, however, full-length music waveforms are normally too long to render a similar large batch size as images. CLMR addressed this by a short fixed audio input of ~2.7s, leading to a trade-off between fair contrastive learning performance and sufficient temporal contextual information.
\cite{wu2021multi} combined multi-task learning with SSL in the pre-training stage, however, it brought marginal improvement because selected pretext tasks like waveform reconstruction and log power spectrum guide the representation learning process more focus on detailed acoustic features than discriminative content.

In this paper, we propose S3T, a self-supervised pre-training method with Swin Transformer for music classification. 
S3T builds upon MoCo~\cite{he2020momentum}, a momentum-based contrastive learning paradigm decoupling the size of negative samples from the batch size with a dynamic dictionary. S3T employs the Swin Transformer~\cite{liu2021Swin} as its backbone on music time-frequency domain. Swin Transformer is a general-purpose vision backbone showing advances in several recognition tasks. In S3T, Swin Transformer helps extract meaningful features from music spectrograms under its inductive biases of locality, hierarchy, and translation invariance. In addition, we design a music augmentation pipeline and two pre-processors to adapt S3T better to music domain.

We evaluate S3T on two music classification tasks, i.e., genre classification and music tagging with linear classifiers. Empirical results show that S3T outperforms CLMR on both tasks, with top-1 genre classification accuracy on the GTZAN dataset being 12.5 percents higher, and the PR-AUC of music tagging on the MagnaTagATune dataset being 4.8 percents higher. S3T also surpasses the task-specific supervised state-of-the-art methods on both tasks. Moreover, S3T demonstrates its label efficiency with using only 10\% labeled data surpassing CLMR trained with full labeled data on both tasks.

\begin{figure*}[!thb]
    \vspace{-0.8cm}
	\centering
	\includegraphics[width=\textwidth,trim={0.5cm 7.5cm 7.5cm 7.5cm}, clip=true]{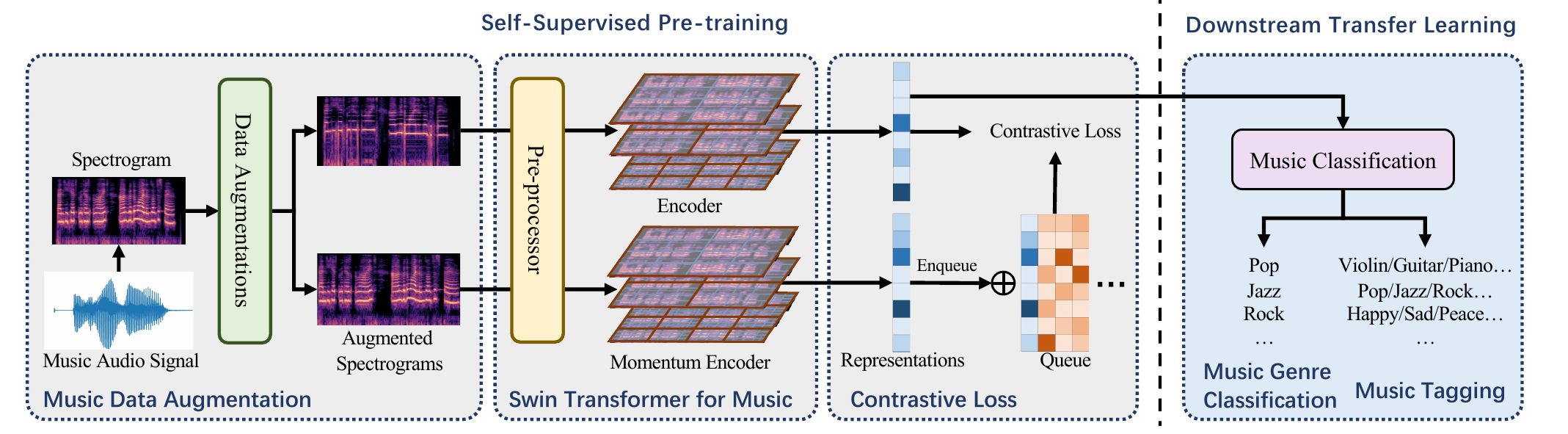}
	\caption{The pipeline of S3T. The left part illustrates the self-supervised pre-training stage and the right part introduces the downstream transfer learning stage.}
	\label{fig:pipeline_overview}
\end{figure*}

		

\section{Method}
\label{sec:method}
In this section, we introduce the design of S3T. We first describe the overall pipeline of S3T and then introduce the major components specially for music audio. 

\subsection{Pipeline Overview}
An overview of S3T is shown in Figure~\ref{fig:pipeline_overview}, which is based on a contrastive learning paradigm MoCo with a siamese structure. S3T contains two stages: self-supervised pre-training and downstream transfer learning. The pre-training stage involves three modules: music data augmentation, encoder and momentum encoder with the same Swin Transformer structure, and a contrastive objective associated with a dynamic dictionary. During pre-training, S3T first converts music audio signal into time-frequency spectrogram and randomly applies a series of music data augmentations to yield two augmented spectrograms, one as query while another as key. S3T then applies a specially designed pre-processor on them and sends them to two encoders respectively. Finally, the encoded query looks up its corresponding key from the dictionary to learn a contrastive objective. In the transfer learning stage, we train simple linear classifiers upon the self-supervised representations for downstream music classification tasks.

\subsection{Self-Supervised Learning Framework}
\label{sec:ssl}
S3T employs a momentum-based contrastive learning framework, MoCo, on the music time-frequency domain. S3T has two encoders, a query encoder $Enc_q$ and a momentum encoder $Enc_k$. Each encoder encodes the query $q=Enc_q(x^q)$ and the key $k^+=Enc_k(x^{k^+})$ respectively. S3T enqueues the current mini-batch encoded keys into a large queue as a look-up dictionary while dequeues the oldest mini-batch. Then the encoded query $q$ looks up its key $k^+$ in the dictionary with InfoNCE~\cite{oord2018representation} as the contrastive loss, whose value is low if $q$ is similar to $k^+$ and dissimilar to all other keys in the dictionary. The dictionary is much larger than a training batch which enables rich negative music samples to perform contrast. Such mechanism mitigates the exhaustive demand of those in-batch contrastive methods like SimCLR as they consider negative samples within one batch. To maintain the dictionary tractable and consistent, the key encoder is updated by a momentum-based moving average of the query encoder instead of back-propagation.

\subsection{Music Data Augmentation}
\label{sec:data_aug}
Rich data augmentations promote contrastive learning by encouraging learning augmentation-invariant representations. To increase the robustness of music representations, we propose a music augmentation pipeline based on \cite{park2019specaugment,al2021clar} adding more complexity and stochasticity. Firstly, apply \textit{Random Multi-crop} on an input sample, yielding 2 crops with random lengths. S3T assumes that two crops from the same music sample have similar representations. Then the other 4 either frequency or temporal transformations are randomly applied on the two crops with specific probability $p$. 
\begin{itemize}[leftmargin=*,itemsep=0pt,topsep=0pt,parsep=0pt]
\item\textbf{Random Multi-crop}
Cut two crops from a length $T$ spectrogram with length $T*r_1$ and $T*r_2$ randomly, $r_1$ and $r_2$ are uniformly sampled from a ratio range [0.1, 0.9].

\item\textbf{Random Frequency Masking}
Mask $N$ frequency segments and each segment has a length of $F$. $N$ and $F$ are uniformly sampled from the range [1, 5] and [5, 30] respectively, with total mask length constrained. Randomly applied with $p=0.5$.

\item\textbf{Random Time Masking}
Mask $N$ temporal segments with each length of $t=T * r$, $N$ and $r$ are sampled from uniform distribution [1, 10] and [0.01, 0.2], total mask length is restricted to $T*0.4$. Randomly applied with $p=0.5$. 

\item\textbf{Time Warping} 
Following \cite{park2019specaugment}, a random point along the time axis passing through the center of spectrogram within the time steps $(W, T-W)$ is warped either to left or right by a distance $W$ sampled from [0, 10] uniformly. Randomly applied with $p=0.4$. 

\item\textbf{Random Shifting} 
Randomly applied pitch shift or time shift with $p=0.4$. Horizontally or vertically shift to either directions $t$ or $f$ steps, both $t$ and $f$ are sampled from a uniform distribution [1, 10].
\end{itemize}

\subsection{Swin Transformer for Music}
\label{sec:swin_for_music}
Swin Transformer~\cite{liu2021Swin} is proposed as a general-purpose vision backbone and shows advances in both supervised and self-supervised works~\cite{xie2021moby,li2021esvit}. Swin Transformer learns a hierarchical representation by gradually merging neighboring small-sized patches as layer goes deeper. Besides, the shifted windowing mechanism of Swin Transformer limits the self-attention computation within non-overlapping windows while also allowing cross-window connections. In S3T, the Swin Transformer extracts multi-resolution time-frequency features with self-attention within hierarchical partitioned spectrograms. Meanwhile, such scheme allows capturing cross-frequency and cross-temporal features. To leverage powerful Swin Transformer while preventing rigidly viewing spectrograms as images, we design two different spectrogram-specific pre-processors, Frequency Tiling and Time Folding, to make proper adaptation to the music domain.
\begin{itemize}[leftmargin=*,itemsep=0pt,topsep=0pt,parsep=0pt]
\item\textbf{Frequency Tiling} 
Replicate a spectrogram with shape of $F \times T$ in frequency domain $n$ times to $(F*n) \times T$, then cut out extra high frequency part to form a shape of $T \times T$.

\item\textbf{Time Folding}
Fold a music spectrogram along time axis ${n}$ times. i.e. a spectrogram with shape $F \times T$ is directly collapsed into $(F * n) \times (T / n)$, with equal length and width.
\end{itemize}

\section{Experiments and Results}
\label{sec:exp}
\subsection{Datasets}
We conducted the self-supervised pre-training of S3T on an in-house dataset of 1 million music recordings covering various genres, languages, and themes. Each music recording is sampled at 16 kHz, with a length ranging from 3 to 5 minutes. The pre-trained model was then transferred to two downstream tasks: music genre classification and music tagging. The genre classification experiments were run on the GTZAN and the FMA datasets, while the music tagging experiments were evaluated using the MagnaTagATune dataset. The details of the three datasets are as follows:
\begin{itemize}[leftmargin=*,itemsep=0pt,topsep=0pt,parsep=0pt]
\item\textbf{GTZAN} \cite{tzanetakis2002musical} contains 1,000 30-seconds-long music clips distributed across 10 distinct genres, with a sampling rate of 22.05 kHz. In this paper, we used the ``fault-filtered" split described in~\cite{kereliuk2015deep} to make a fair comparison with \cite{spijkervet2021contrastive,castellon2021codified}.

\item \textbf{FMA} \cite{defferrard2017fma} is a music analysis dataset collected from Free Music Achieve. We used the small version of FMA, which consists of 8,000 music clips of 8 genres, with each clip being around 30 seconds long and sampled at 44.1 kHz. We followed the official 8:1:1 split to divide FMA into train/valid/test subsets.

\item \textbf{MagnaTagATune} (MTAT) \cite{law2009evaluation} contains 25,863 29-seconds-long music clips extracted from 5,223 songs, 445 albums and 230 artists. Each clip is associated with a vector of binary annotations of 188 tags. We followed the official partition using the top 50 most popular tags for evaluation.
\end{itemize}

\subsection{Experimental Setup}
\label{sec:setup}
\noindent\textbf{Model Input} Each music sample input to S3T was first resampled to a sampling rate of 22.05 kHz and then converted to a 84-dim constant-Q transform (CQT) spectrogram, with the number of bins per octave as 12, the hop size as 512, and the Hann window as the windowing function. To improve the training efficiency, we further compressed the CQT spectrogram by a factor of 100 along the time axis. The empirical results demonstrate that though compressed 100 times S3T can still learn meaningful representations from CQTs yielding competitive performance on music classification.


\noindent\textbf{Feature Extractor} We followed the setting of Swin-T, the tiny version of Swin Transformer, with hidden channel number of 96, \{2, 2, 6, 2\} layers in each block respectively, and a drop path rate of 0.1. To adapt the 84-dim CQT-spectrogram to be used for Swin-T, we changed the model input size to 256 $\times$ 256 with a window size 8. We followed the MoCo v2 configuration using a projection head with two hidden fully-connected layers to project the raw representation to a 128-dim hidden space. Unless otherwise mentioned, we used this setup in all experiments. 
As for linear evaluation, we followed common linear protocol using a 1-layer MLP with representations as input features for each downstream task.

\noindent\textbf{Pre-training Setup} Follow \cite{he2020momentum}, the total size of the queue is 65,536. The pre-training of S3T was run using 4 Tesla V100 GPUs, with a batch size of 128 on each GPU and 400 training epochs in total.
Following~\cite{liu2021Swin}, we employed the AdamW~\cite{loshchilov2017decoupled} optimizer with default setting, with a cosine decay learning rate scheduler and 20 epochs of linear warm-up. 
The initial learning rate was set to 0.0005.

\noindent\textbf{Downstream Setup} During the transfer learning stage, we trained a simple linear classifier for each downstream task by freezing the feature extractor. The training of each downstream task was run using 1 Tesla V100 GPU with 50 epochs in total. A batch size of 64 was used for GTZAN and 256 for both FMA and MTAT. 
We employed the AdamW optimizer with an initial learning rate of 0.001 and a weight decay of 0.05. 
Besides, we used the cosine decay  learning rate scheduler with a linear warm-up for 5 epochs. 

\subsection{Experimental Results and Analyses}
\subsubsection{Main Results}
We evaluated S3T on both music genre classification and music tagging, and compared our results with other methods using a linear protocol as described in Section~\ref{sec:setup}.: 
1) \textit{Random Init}, which trains a linear classifier with a randomly initialized feature extractor;
2) \textit{Supervised}, which pre-trains the feature extractor with labeled data in a supervised manner; 
3) \textit{SSL~(CLMR)}, which pre-trains the feature extractor with unpaired music by contrastive learning.  For comparison, we use average tag-wise area under the receiver operating characteristic curve (ROC-AUC) and average precision (PR-AUC) scores as metrics for music tagging on MTAT, and use top-1 accuracy for genre classification on GTZAN and FMA. 


\begin{table}[t]
\centering
\small 
\vspace{-0.3cm}
\begin{tabular}{ c | c c c c}
	\toprule
	\multirow{2}{*}[-0.7ex]{Method}  & \multicolumn{2}{c}{MTAT} & FMA & GTZAN \\
    \cmidrule(lr){2-3} 
    \cmidrule(lr){4-4} \cmidrule(lr){5-5}
	& PR-AUC & ROC-AUC  & Top-1(\%) & Top-1(\%) \\
	\midrule
	\textit{Random Init} & 18.0 & 73.5 & 26.0 & 27.6 \\
    \textit{Supervised} & 38.3 & \textbf{90.6} & 42.7 & 79.0 \\
    \textit{SSL (CLMR)} & 36.1 & 89.4 & 48.4 & 68.6 \\
    \textit{S3T (Ours)} & \textbf{40.9} & 89.9 & \textbf{56.4} & \textbf{81.1} \\
	\bottomrule
\end{tabular}
\vspace{-0.2cm}
\caption{Comparison of linear evaluation results of S3T with other three methods. Supervised results are collected from~\cite{castellon2021codified,tran2020attention}.}
\label{tab:main_results}
\end{table}


S3T outperforms \textit{Random Init}, \textit{CLMR}, and respective \textit{Supervised} methods on all tasks, as shown in Table~\ref{tab:main_results}. Especially, for music tagging, S3T achieves a PR-AUC of 40.9\% and a ROC-AUC of 89.9\% on MTAT, exceeding the \textit{CLMR} by 4.8 percents and the \textit{Supervised} by 2.6 percents on PR-AUC. For music genre classification, S3T surpasses \textit{CLMR} by 12.5 percents and 8.0 percents on top-1 accuracy on GTZAN and FMA respectively and exceeds the \textit{Supervised} pre-trained method by 13.7 percents on FMA and 2.1 percents on GTZAN.
Experimental results demonstrate that S3T can make good use of massive unlabeled music and pre-train a feature extractor with good representation learning ability for music classification tasks.

\begin{table}[t]
\centering
\small 
\begin{tabular}{ c | c c c c }
	\toprule
	\multirow{2}{*}[-0.7ex]{Setting} & \multicolumn{2}{c}{MTAT} & \multicolumn{2}{c}{FMA} \\
	\cmidrule(lr){2-3}\cmidrule(lr){4-5}
	& PR-AUC & ROC-AUC & Top-1(\%) & Top-5(\%) \\
	\midrule
	\textit{CLMR} & \underline{36.1} & 89.4 & \underline{48.4$^*$} & \underline{90.6$^*$}\\
	\textit{Supervised} & 38.3 & 90.6 & 42.7 & -- \\
	\midrule
	\textit{S3T-1\%} & 27.2 & 77.4 & 39.0 & 84.8 \\
	\textit{S3T-10\%} & \underline{37.2} & 87.8 & \underline{50.6} & \underline{91.9} \\
	\textit{S3T-100\%} & 40.9 & 89.9 & 56.4 & 95.3 \\
	\bottomrule
\end{tabular}
\vspace{-0.2cm}
\caption{Comparison of linear evaluation results with different proportions of labeled data on MTAT for music tagging and FMA for genre classification. \textit{$^*$ indicates reproduced results with official release of 10,000-epoch model.}}
\label{tab:label_efficency}
\vspace{-0.3cm}
\end{table}

\subsubsection{Method Analysis}
In this part, we analyze the label efficiency, feature extractor choice, and pre-processor choice of S3T.

\noindent\textbf{Label Efficiency} To study the effectiveness of S3T, We evaluated S3T with limited labeled subsets (1\%, 10\%) of MTAT and FMA for two classification tasks and report the results in Table~\ref{tab:label_efficency}.
Noticeably, for the music tagging task, S3T with only 10\% labeled data achieves a PR-AUC 37.2\% on MTAT, which exceeds \textit{CLMR} with full labeled data (36.1\%), and performs competitively with \textit{Supervised} pre-trained result with full labeled data (38.3\%). For music genre classification, S3T with only 10\% labeled data outperforms both \textit{CLMR} and \textit{Supervised} pre-trained method with full labeled data on FMA by 2.2 percents and 7.9 percents. Evaluation results on both tasks demonstrate S3T has strong representation learning capability even with only 10\% or 1\% labeled data.

\begin{table}[t]
\centering
\setlength\tabcolsep{20pt}
\small 
\vspace{-0.3cm}
\begin{tabular}{ c | c  c }
	\toprule
	\multirow{2}{*}[-0.7ex]{Backbone} & \multicolumn{2}{c}{GTZAN} \\
	\cmidrule(lr){2-3}
	 & Top-1(\%) & Top-5(\%)\\
	\midrule
	\textit{ResNet50} & 73.6 & 93.4 \\
	\textit{MSNet-BiGRU} & 64.5 & 91.9 \\
    \textit{Swin-T} & \textbf{81.1} & \textbf{97.2} \\
	\bottomrule
\end{tabular}
\vspace{-0.2cm}
\caption{Comparison of the results on GTZAN among different backbones.}
\label{tab:backbone}
\end{table}

\noindent\textbf{Feature Extractor}
To investigate the effect of model backbone on the performance of music classification, we compared the performance of S3T on genre classification when using \textit{ResNet50}~\cite{he2016deep}, \textit{MSNet}~\cite{zhu2021msnet}\textit{-based BiGRU}~\cite{cho2014learning}, and \textit{Swin-T} as the backbone and report the results in Table~\ref{tab:backbone}. The results indicate that Swin-T has superior performance than both CNN-based and BiGRU-based backbones.


\noindent\textbf{Pre-processor}
Our last experiment is to compare two data pre-processors described in Section~\ref{sec:swin_for_music} on music genre classification, and the results are listed in Table~\ref{tab:pre_processor}. As shown in the table, \textit{Time Folding} outperforms \textit{Frequency Tiling} slightly, mainly because it remains more original contextual information instead of just repeating small segments of input sample.

\begin{table}[t]
\centering
\setlength\tabcolsep{20pt}
\small 
\begin{tabular}{ c | c  c }
	\toprule
	\multirow{2}{*}[-0.7ex]{Pre-processor} & \multicolumn{2}{c}{GTZAN} \\
	\cmidrule(lr){2-3}
	& Top-1(\%) & Top-5(\%) \\
	\midrule
	\textit{Frequency Tiling} & 78.3 & 96.6 \\
    \textit{Time Folding} & \textbf{81.1} & \textbf{97.2} \\
	\bottomrule
\end{tabular}
\vspace{-0.2cm}
\caption{Comparison of the results on GTZAN between two pre-processors.}
\label{tab:pre_processor}
\vspace{-0.5cm}
\end{table}

\section{Conclusion}
\label{sec:conclusion}
In this paper, we propose S3T, a self-supervised pre-training method with Swin Transformer for music classification, which leverages massive unlabeled music data to improve the performance of music classification and reduce the dependence on a considerable amount of labeled music data. S3T is built upon a momentum contrast paradigm and firstly applies the Swin Transformer as a feature extractor for music classification. Besides, S3T contributes a music data augmentation pipeline and two specially designed pre-processors. Experiments show that S3T outperforms previous self-supervised methods and becomes competitive to task-specific supervised pre-trained method. Our further method analysis verifies the effectiveness of detailed designs in S3T, especially its label efficiency. In the future, we will consider pre-training on a larger amount of unlabeled data, and explore a feature extractor better leveraging temporal contextual information of music, and combine S3T with a specially designed classifier for music classification.

\bibliographystyle{style/IEEEbib}
\bibliography{bib/refs}

\end{document}